\documentclass[aps,pra,preprint,showkeys]{revtex4-2}
\usepackage[T1]{fontenc}
\usepackage[utf8]{inputenc} 
\usepackage{graphicx}
\usepackage{amsmath}
\usepackage{amsfonts}
\usepackage{subfigure}
\usepackage{xcolor}
\usepackage{mathptmx}
\usepackage{epstopdf}
\usepackage{braket}
\usepackage{bm}

\newmuskip\pFqmuskip
\newcommand*\pFq[6][8]{%
	\begingroup 
	\pFqmuskip=#1mu\relax
	\mathcode`\,=\string"8000
	\begingroup\lccode`\~=`\,
	\lowercase{\endgroup\let~}\pFqcomma
	{}_{#2}F_{#3}{\left[\genfrac..{0pt}{}{#4}{#5};#6\right]}%
	\endgroup
}
\newcommand{\pFqcomma}{\mskip\pFqmuskip}

\begin{document}

\title{Comment on "Continuous simultaneous measurement of position and momentum of a particle"}
   
	\author{Adelcio C. Oliveira} 
	\email{adelcio@ufsj.edu.br}
	\address{Departamento de F\'isica e Matem\'atica, Universidade Federal de S\~ao
		Jo\~ao Del Rei, C.P. 131, Ouro Branco, MG, 36420 000, Brazil }
	\begin{abstract}
		In a recent paper, [Gampel, F. and Gajda, M., Phys. Rev. A 107, 012420, (2023)], the authors claimed they are proposing a new model to explain the existence of classical trajectories in the quantum domain.
The idea is based on simultaneous position and momentum measurements and a "jump Markov process".
Consequently, they have interpreted the emergence of classical trajectories as sets of detection events.
They successfully implemented the model for a free particle and for one under a harmonic potential. Here,
we show that the continuous observation limit is a realization of a coherent semiclassical expansion; Also, as has already been demonstrated, the jump process is not necessary and is not observable. In other words,
the collapse, as they propose, is a non-go theorem; even if it is real, it can not be measured under the needed
assumptions to obtain Newtonian classical dynamics.

	\end{abstract}

\maketitle
\begin{figure}[!ht]
\centering
\includegraphics[width=12cm,height=6cm,keepaspectratio]{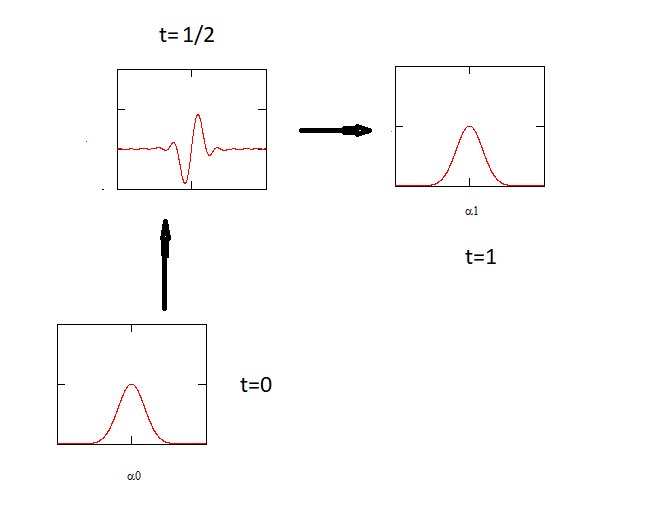}
\caption{Schematic representation of the measuring process .}
\label{Med}
\end{figure}

The basic idea is to consider a single particle system monitored in high frequency. This monitoring process is realized in a way that the position and momentum are observed simultaneously and independent process, it can be realized beyond the standard uncertain principle \cite{Ozawa}. This procedure is not only a scientific advance but has been used in some practical applications such as metrology \cite{Tsarev,Cimini}. The uncertain principle does not limit the measurement precision, but it limits the minimum uncertainty that a quantum state can have, or in other words, it gives an additional restriction on   $\mathcal{L}^2$ set for states that can be observed in the quantum domain.  After a highly accurate simultaneous position and momentum measurement (CMPM), the quantum state can not have a thickness lower than uncertain principle demands, thus the narrowest that it can be is a coherent state, this approach was first investigated in  \cite{Oliveira2014} and in reference \cite{Oliveira2021}. It was also shown that CMPM can be implemented by the continuous reset approach. Now we show the schematic representation of the measurement process in figure \ref{Med}, where at time $t=0$  the state is a coherent state centered at $\alpha_0$, and after the measurement in time $t=1$ it is a coherent state centered at $\alpha_1$, in the middle time it evolves under its Hamiltonian. Gampel and Gajda \cite{Gampel} claim that the detector would produce a collapse of the wave-function in some $\ket{\alpha}$ where $\alpha=x_m+ip_m$, where $(x_m,p_m)$ is given by the measurement, the same approach of \cite{Oliveira2014,Oliveira2021}. In addition to that, they assume that the result of the measurement will have some uncertain and the state in $t=1$ will be a state in the set of coherent states $\{\ket{\alpha_r}\}$ where $\alpha_r$ is in the measurement precision range, thus, this randomness of the final state in time $t=1$ is responsible for the jump process. At that point, we observe that if the system was measured in high precision momentum and position procedure \cite{Ozawa} then this "jump" is inner the precision of measurement, then it will not be observed. For any practical reasons, a stochastic approach would demand many repetitions of the same setup and a mean procedure, then the measured pair $(x_m,p_m)$ will converge to its mean. In the reference \cite{Oliveira2014} we have shown that the dynamics converge to the classical Newtonian dynamics in the continuous observation limit.

\section*{Numerical Approach}

To illustrate this fact, we numerically implemented the "jump approach" for the quartic oscillator, its Hamiltonian is given by
\begin{equation}
    \widehat{H}=\hbar \omega \widehat{a}^\dagger\widehat{a}+\hbar^2\lambda(\widehat{a}^\dagger\widehat{a})^2
\end{equation}
where $\widehat{a}^\dagger$ and $\widehat{a}$ are the usual creation and annihilation operators.

\begin{figure}[!ht]
\centering
\includegraphics[width=15cm,height=8cm,keepaspectratio]{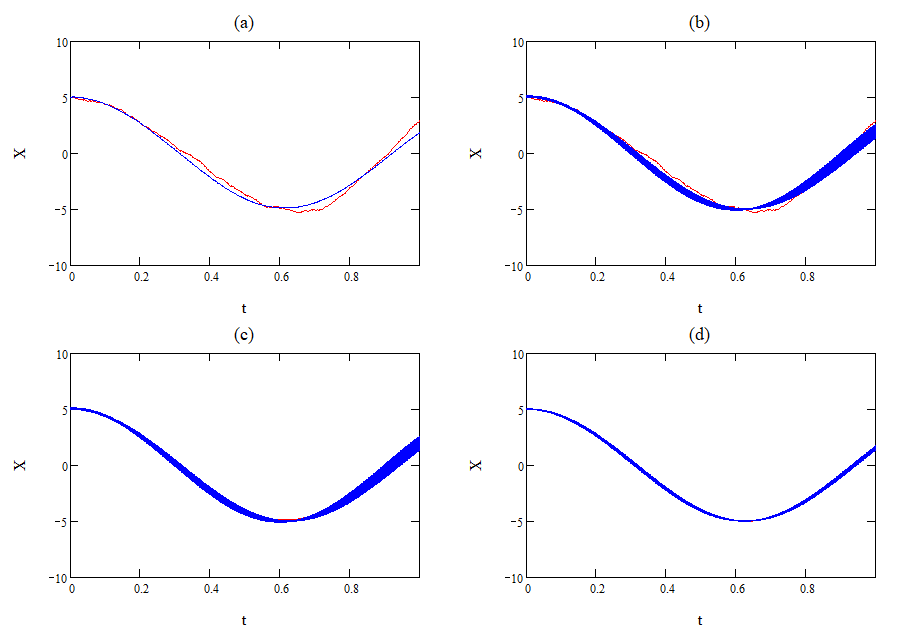}
\caption{(a) Quantum expectation value of the position operator $X$ under a continuous measurement and "jump" approach. In red is shown one realization, and in blue is the mean of 100 realizations. (b) In red same as the red curve in (a) and in blue is the error bar of classical dynamics. (c) Same as (b) but for the quantum mean of 100 realizations of the process. In those figures, the uncertainty in the $\alpha$  is $\delta \alpha=0.1$.  (d) Same as (b) for lower measurement precision, $\delta \alpha=0.01$. The other parameters are: $\hbar=1$, $\omega=0$, $\lambda=1$ and $\alpha_0=5$.}
\label{Jump}
\end{figure}
\begin{figure}[!ht]
\centering
\includegraphics[width=6cm,height=8cm,keepaspectratio]{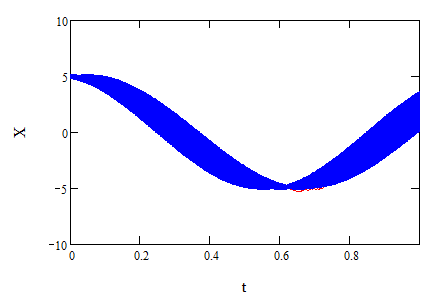}
\caption{ Same as figure \ref{Jump}-b including an imprecision on time $\delta t =0.05$.}
\label{Jump2}
\end{figure}

In figure \ref{Jump} we show the results for the quantum position expectation value under a continuous monitoring approach and "jump" (CMPMJ), the model with collapse proposed by Gampel and Gajda. In (a) red is one realization and blue is the mean of 100 realizations. As we can see the abrupt changes in the expectation value that disappear in the mean procedure. In (b) we show the one realization expectation model and the error bar of the classical equivalent process. It is clear that the curve is in the inner of the region in almost of the time interval. On the other hand, in (c) we plot the mean, which is clearly inside the error bar. As we commented above and in the reference \cite{Oliveira2014} the classical Newtonian dynamics are reproduced only under high-precision measurement, or mathematically,  for the quartic model, it means that $\frac{\delta \alpha}{|\alpha|}\ll 1$, that is the case of \ref{Jump}-d, where $\frac{\delta \alpha}{|\alpha|}= 2 \times 10^{-3}$,  and the one realization curve is completely inside the error bar region. Another important point is when there is a small imprecision in time measurement, it erasures most of the discrepancies observed in fig \ref{Jump}-b, as was shown in reference \cite{Oliveira2012}.
\\ In the cited paper \cite{Gampel}, the authors claim that the open system approach leads to statistical prescriptions that converge to the classical Liouville master equations, which is in part true it is necessary a combination of factors, large action, experimental limitations and interaction with the environment \cite{Oliveira2012}. Not less important, as pointed out by Angelo ‘‘inevitable conception that Physics must ferment a
realistic description of only one system.’’ Indeed, ‘‘nature as a whole may be thought of as an individual system existing
only one time, with no need for repetitions and not as an ‘ensemble of systems’’ see \cite{Angelo}. In the "classical world," the repetition of the experiment can not change the results, i.e. the same initial conditions would produce the same final result,  the statistics matter when we have experimental limitations and the initial conditions or the Hamiltonian are only partially known, but the repetition of the procedure with some variations in the parameters is not a sufficient condition to obtain a classical statistical prediction of Liouville type as was shown \cite{Oliveira2012}. Again, if we assume that the system is under CMPMJ scheme \cite{Ozawa} then the experimental limitations are not significant and the measure does not drive it into a classical Liouville dynamics as observed in figure \ref{Jump}-b, the result is shown on figure \ref{Jump2}, note that even for small imprecision on time measurement $\delta t=0.05$ enlarge significantly the error bar, and the quantum dynamics is inside the error bar in almost everywhere. Since the discontinuities in the trajectory are not observable, then for any practical reason CMPMJ is indistinguishable from CMPM and therefore we need not model this virtual collapse.  

\section*{Analytical Approach}

In order to obtain an analytical approach, we observe that for a general Hamiltonian $\widehat{H}$, and initial state $\widehat{\rho}(0)$ at time $t$ we can write \cite{Oliveira2003,Romero2004,Reis} its time evolution as  
\begin{eqnarray}
\widehat{\rho }(t) &=&\widehat{U}_{sc}(t)\{\widehat{\rho }(0)+\dfrac{i}{%
\hbar }\int_{0}^{t}[\widehat{\rho }(0),\widehat{\Delta }_{s1}]dt_{1}+\left(
\dfrac{i}{\hbar }\right) ^{2}\int_{0}^{t}\int_{0}^{t_{1}}[[\widehat{\rho }%
(0),\widehat{\Delta }_{s1}],\widehat{\Delta }_{s1}]dt_{2}dt_{1}  \label{1.32}
\\
&&+\left( \dfrac{i}{\hbar }\right)
^{3}\int_{0}^{t}\int_{0}^{t_{1}}\int_{0}^{t_{2}}[[[\widehat{\rho }(0),%
\widehat{\Delta }_{s3}],\widehat{\Delta }_{s2}],\widehat{\Delta }%
_{s1}]dt_{3}dt_{2}dt_{1}+...{\LARGE \}}\widehat{U}_{sc}^{\dagger }(t)  \notag
\end{eqnarray}%
The operators $\Delta_i$ and $\widehat{U}_{sc}$ are defined by the
perturbation $\delta $, that is defined as%
\begin{equation}
\delta (t)=\widehat{H}-\widehat{H}_{sc}(t)  \label{delta}
\end{equation}%
thus we have

\begin{equation*}
\left\{
\begin{array}{c}
\widehat{H}=\widehat{H}_{sc}(t)+\widehat{\delta }(t), \\
i\hbar \frac{d}{dt}\widehat{\rho }=-[\widehat{\rho },\widehat{H}_{sc}]-[%
\widehat{\rho },\widehat{\delta }(t)].%
\end{array}%
\right.
\end{equation*}
In the interaction picture, in terms of $\widehat{H}_{sc},$ we define
\begin{equation}
\widehat{\rho }_{I}=\widehat{U}_{sc}^{\dagger }(t)\widehat{\rho }\widehat{U}%
_{sc}(t)
\end{equation}%
where $\widehat{U}_{sc}(t)$ is the semiclassical time evolution operator,
then we have%
\begin{equation}
i\hbar \frac{d}{dt}\widehat{\rho }_{I}=-[\widehat{\rho }_{I},\widehat{\Delta
}_{s}].  \label{1.31}
\end{equation}%
where \bigskip
\begin{equation}
\widehat{\Delta }_{s}(t)=\widehat{U}_{sc}^{\dagger }(t)\widehat{\delta }(t)%
\widehat{U}_{sc}(t),  \label{Delta}
\end{equation}%
and $\widehat{U}_{sc}(t)$ is defined as
\begin{equation}
i\hbar \frac{\partial }{\partial t}\widehat{U}_{sc}(t)=\widehat{H}_{sc}(t)%
\widehat{U}_{sc}(t).
\end{equation}%

\ If the classical hamiltonian can be writeen as
\begin{equation*}
H_{\text{cls}}=\hbar \omega \,\alpha ^{\ast }\alpha +\sum_{m,n}A_{m,n}\left(
\alpha ^{\ast }\right) ^{m}\alpha ^{n},
\end{equation*}%
then, the semiclassical Hamiltonian, for one degree of freedom \cite%
{Oliveira2003}, is

\begin{equation}
\begin{split}
\widehat{H}_{\text{sc}}(\alpha (t))& =\hbar \omega \,\widehat{a}^{\dagger }%
\widehat{a}+\sum_{m\neq 0}m\,A_{m,m}\left( \alpha ^{\ast }\right)
^{m-1}\alpha ^{m-1}(\widehat{a}^{\dagger }\widehat{a}-\alpha ^{\ast }\alpha
)+\sum_{m,n}A_{m,n}\left( \alpha ^{\ast }\right) ^{m}\alpha ^{n} \\
& \qquad +\sum_{m\neq n}m\,A_{m,n}\left( \alpha ^{\ast }\right) ^{m-1}\alpha
^{n}(\widehat{a}^{\dagger }-\alpha ^{\ast })+\sum_{m,n}n\,A_{m,n}\left(
\alpha ^{\ast }\right) ^{m}\alpha ^{n-1}(\widehat{a}-\alpha ).
\end{split}
\label{9}
\end{equation}
In the continuous simultaneous position and momentum measurements the corrections in the approximation vanish and we obtain

\begin{equation}
   \lim_{n \longrightarrow \infty}\widehat{\rho }(t) =\widehat{U}_{sc}(t)\widehat{\rho }(0)\widehat{U}_{sc}(t)^{\dagger}, 
\end{equation}
where $n$ is the number of measurements in the time interval, see reference \cite{Oliveira2014} for the details. The state is a Gaussian and its centroid obeys Newtonian dynamics \cite{Oliveira2003}. We emphasized that it is an analytical result that is valid for any Hamiltonian, for $SU2$ algebra see \cite{Romero2004,Reis}. As was numerically shown in \cite{Oliveira2014,Oliveira2021}, it is not necessary to have continuous monitoring, if the time interval between two measurements is small, then  

\begin{equation}
   \widehat{\rho }(t)  \approx \widehat{U}_{sc}(t)\widehat{\rho }(0)\widehat{U}_{sc}(t)^{\dagger}, 
\end{equation}
and for any practical reason the Newtonian classical limit is achieved. That is valid for any general Hamiltonian, not restricted to chaotic dynamics as referred by \cite{Gampel}.
In figure [2] of reference \cite{Gampel} they implemented the "jump" simulation for a harmonic potential, for this potential $\widehat{\rho }(t) = \widehat{U}_{sc}(t)\widehat{\rho }(0)\widehat{U}_{sc}(t)^{\dagger}=\ket{\alpha(t)}\bra{\alpha(t)
} $, where $\ket{\alpha(t)}$ is a coherent state and the centroid obeys the classical equations of motion, that is exactly the reason why the coherent states belong to the set known as quasi-classical states \cite{Klauder}.

Quantum Darwinism \cite{Zurek} predicts that the environment and system interaction generates a proliferation of quantum states, known as pointer states, as well as shifting the probabilistic balance towards to states that are termed classical. The CMPM can then be understood as an additional selection rule in Quantum Darwinism, when observables that do not commute are measured simultaneously by independent processes, the apparatus acts on the system selecting only states that presents minimum variances of the observables. Understanding how this process takes place is the subject of future investigation.

\section*{Final remarks}
In summary, the classical Newtonian limit can be achieved by high precision simultaneous measurement in an almost continuous process as stated in \cite{Oliveira2014, Gampel,Oliveira2021}, but the procedure of adopting a non-continuous process of collapse as proposed by \cite{Gampel} is not necessary and not an observable process. Instead of that, the semiclassical approximation gives the exact result in the continuous monitoring limit \cite{Oliveira2014,Oliveira2021} and is a good approximation for the not exactly continuous measurement process.

	\section*{Acknowledgments}
    The author acknowledges M. M. Oliveira for his helpful comments and suggestions.   
	The author also acknowledges the support of the Brazilian Agency Funda\c{c}\~{a}o de Amparo \`{a} Pesquisa do Estado de Minas Gerais
	(FAPEMIG). and the support provided by the Brazilian Agency Coordena\c{c}\~{a}o de Aperfei\c{c}oamento de Pessoal de N\'{i}vel Superior (CAPES) and the support of the National Council for Scientific and Technological Development – CNPq.


\begin{thebibliography}{99}

\bibitem{Ozawa} M. Ozawa, Phys. Rev. A 67 (2003) 042105.

\bibitem{Tsarev} D. V. Tsarev, S. M. Arakelian, You-Lin Chuang, Ray-Kuang Lee, and A. P. Alodjants, "Quantum metrology beyond Heisenberg limit with entangled matter wave solitons," Opt. Express 26, 19583-19595 (2018)

\bibitem{Cimini} Cimini, V., Polino, E., Belliardo, F. et al. Experimental metrology beyond the standard quantum limit for a wide resources range. npj Quantum Inf 9, 20 (2023).

\bibitem{Oliveira2014} Oliveira, A. C.. Classical Limit of Quantum Mechanics Induced by Continuous Measurements.  Physica. A,  V. 392, p. 655-668, 2014. 

\bibitem{Oliveira2021} Oliveira, A. C.,
Continuum reset dynamics as a pathway to Newtonian classical limit of Quantum Mechanics,
Physica A, V. 579, 126099,  2021.

\bibitem{Gampel}  Gampel, F. and Gajda, M.,
Phys. Rev. A 107, 012420, (2023).



\bibitem{Oliveira2012} Oliveira, A. C., Bosco de Magalhães, A. R. and Peixoto Faria,  J.G., Physica A 391 (2012) 5082.


\bibitem{Angelo} R.M. Angelo, Low-resolution measurements induced classicality, 2008. arXiv:0809.4616.


\bibitem{Oliveira2003} Oliveira, A. C., Nemes, M. C.,Fonseca Romero, K. M., Phys. Rev. E 68 (2003) 036214.

\bibitem{Romero2004} Fonseca Romero, K.M., Nemes, M. C., Peixoto de Faria, J.G., and de Toledo Piza,  A.F.R., Phys. Lett. A 327 (2004) 129.

\bibitem{Reis} Reis, M., Nemes, M.C. , Peixoto de Faria, J.G. , Phys. Rev. E 78 (2008) 036220.

\bibitem{Klauder} Klauder, J.R. and Skagerstam,  B., Coherent States, World Scientific, Singapore, 1985.

\bibitem{Zurek} Zurek, W. Quantum Darwinism. Nature Phys 5, 181–188 (2009).

\end{thebibliography}
\end{document}